# Exploring pedagogical content knowledge of physics instructors using the force concept inventory

Alexandru Maries, and Chandralekha Singh

View Online   Export Citation

### ARTICLES YOU MAY BE INTERESTED IN

Teaching introductory undergraduate physics courses using multimedia resources
AIP Conference Proceedings **2109**, 120003 (2019); https://doi.org/10.1063/1.5110147

The MicroBooNE experiment
AIP Conference Proceedings **2109**, 110005 (2019); https://doi.org/10.1063/1.5110144

Gender bias in physics: An international forum
AIP Conference Proceedings **2109**, 030007 (2019); https://doi.org/10.1063/1.5110069

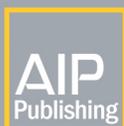
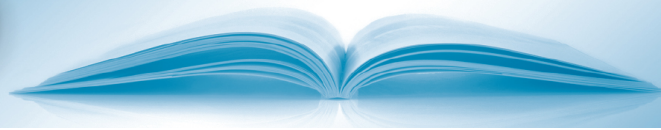

# Exploring Pedagogical Content Knowledge of Physics Instructors Using the Force Concept Inventory

Alexandru Maries[1, a)] and Chandralekha Singh[2]

[1]*Department of Physics, University of Cincinnati, Cincinnati, OH 45221, USA*
[2]*Department of Physics and Astronomy, University of Pittsburgh, Pittsburgh, PA 15260, USA*

[a)]mariesau@ucmail.uc.edu

**Abstract.** The Force Concept Inventory (FCI) has been widely used to assess student understanding of introductory mechanics concepts by educators and physics education researchers. Many of the items on the FCI have strong distractor choices corresponding to students' alternate conceptions in mechanics. Instruction is unlikely to be effective if instructors do not explicitly take into account students' initial knowledge state in their instructional design. We discuss research involving the FCI to evaluate the pedagogical content knowledge of instructors of varying teaching experience. For each item on the FCI, instructors were asked to identify the most common incorrect answer choice of introductory physics students. We also discussed the responses individually with some instructors. Then we used the FCI pre-test and post-test data from a large population (~900) of introductory students to assess the pedagogical content knowledge of the physics instructors related to the FCI. While the physics instructors, on average, performed better than random guessing at identifying introductory students' difficulties with FCI content, they did not identify many common difficulties of introductory physics students. Moreover, the ability to correctly identify students' difficulties was not correlated with the teaching experience of instructors.

## INTRODUCTION

The Force Concept Inventory (FCI) is a 30-question multiple choice survey developed in 1992 by Hestenes, Wells and Swackhammer [1] after early observations [2, 3] that many students enter and leave physics classes with alternate conceptions inconsistent with the scientifically accepted concepts taught in physics classes. This research is part of several comprehensive studies [4–6] that used conceptual assessments such as the FCI and the Test of Understanding Graphs in Kinematics [7] to explore the pedagogical content knowledge (PCK) of instructors of varying teaching experience and teaching assistants. The PCK, as defined by Shulman [8], includes, "understanding of the conceptions and preconceptions that students bring with them to the learning of those most frequently taught topics and lessons." According to this definition, knowledge of common student alternate conceptions is one aspect of PCK [8]. In this study, we explore the PCK of physics instructors with regard to their knowledge of introductory physics students' alternate conceptions related to force and motion, in particular, whether instructors are able to identify the common alternate conceptions of introductory students on individual items in the FCI. Knowledge of the alternate conceptions inconsistent with the scientifically accepted way of reasoning about the concepts can be helpful in devising curricula and pedagogical strategies to improve student understanding. Much physics education research has been devoted to devising and assessing such strategies to help students develop a solid grasp of physics.

The focus is on two research questions related to the PCK of physics instructors at a large research university in the United States.

First, does recent teaching experience influence the ability of the instructor to identify introductory physics students' alternate conceptions in the FCI? We compared the PCK of instructors who had recently taught introductory mechanics (either algebra-based or calculus-based physics) with those who had not.

Second, to what extent do instructors identify specific alternate conceptions of introductory physics students? Is their ability to identify these alternate conceptions context dependent? We identified particular alternate conceptions (e.g., constant force implies constant velocity) in different FCI questions and analyzed instructors' PCK performance in identifying them.

## METHODOLOGY

Thirty physics instructors at a large research university in the United States were given the FCI. For each question, they were asked to identify which one of the four *incorrect* answer choices, in their view, would be most commonly selected by introductory physics students after instruction in relevant concepts if the students did not know the correct answers. Instructors' teaching experience in introductory mechanics varied widely: some were relatively new, some were emeritus professors, and yet others taught introductory mechanics on a regular basis. We compare the average FCI-related PCK performance of the instructors who had taught introductory mechanics courses in the past seven years with those who had not. After the instructors completed the task, we discussed the reasoning for their responses individually with some of them, especially for the questions in which the reasoning was not explicitly provided in the answer choices themselves.

In order to compare the FCI-related PCK performance of different groups of physics instructors, scores were assigned to each instructor. An instructor who selected a particular incorrect answer choice as the most common for a particular question received a PCK score equal to the fraction of introductory students who selected that particular incorrect answer choice. If an instructor selected the correct answer choice as the most common incorrect answer of introductory students (which was rare but happened for a few questions), he or she was assigned a score of zero because the question was to indicate the *incorrect* answer choice. For example, for FCI question 2, the fractions of algebra-based introductory students who selected A, B, C, D, and E were 0.44, 0.25, 0.06, 0.21, and 0.04, respectively. Answer choice A is correct; thus, the PCK score assigned to instructors for each answer choice if they selected it as the most common would be 0, 0.25, 0.06, 0.21, and 0.04 (A, B, C, D, and E, respectively). The total PCK score for an instructor on the task for the entire FCI can be obtained by summing over their scores for all 30 FCI questions.

Our approach to determining the PCK score related to the FCI appropriately weighs the responses of instructors by the fraction of introductory students who selected a particular incorrect response (although the instructors were asked to select the most common incorrect answer choice for each question). The total PCK score can be calculated mathematically for the instructors by defining indices $i$, $j$, and $k$ as follows:
- $i$ = index of instructor (values from 1 to 30)
- $j$ = FCI question number (values from 1 to 30)
- $k$ = incorrect answer choice number for each question (4 incorrect answer choices, values from 1 to 4).

$F_{jk}$ is the fraction of introductory physics students who selected incorrect answer choice $k$ on item $j$ (e.g., $F_{21}$ = 0.25, $F_{22}$ = 0.06, $F_{23}$ = 0.21, $F_{24}$ = 0.04). $I_{ijk}$ corresponds to whether instructor $i$ selected incorrect answer choice $k$ on item $j$ (for a given $i$ and $j$, $I_{ijk}$ = 1 only for the incorrect answer choice $k$, selected by instructor $i$ on item $j$, otherwise $I_{ijk}$ = 0). Thus, the PCK score of the $i$th instructor on item $j$ (referred to as $I_{ij}$) is $I_{ij} = \sum_{k=1}^{4} I_{ijk} \cdot F_{jk}$ $I_{ij} = \sum_{k=1}^{4}(I_{ijk} \cdot P_{jk})$. Then, the total PCK score of the $i$th instructor ($I_i$) on the whole survey can be obtained by summing over all of the questions: $I_i = \sum_{j=1}^{30} I_{ij} = \sum_{j=1}^{30}\left[\sum_{k=1}^{4} I_{ijk} \cdot F_{jk}\right]$. Also, the PCK score of all of the instructors on item $j$ (referred to as $\overline{I_j}$) can be obtained by summing over the instructors: $\overline{I_j} = \sum_{i=1}^{30} I_{ij} = \sum_{i=1}^{30}\left[\sum_{k=1}^{4} I_{ijk} \cdot F_{jk}\right]$.

## RESULTS

Many instructors explicitly noted that the task of thinking from an introductory student's point of view was very challenging; some even confessed that they did not feel confident about their performance in identifying the most common incorrect answers.

### Influence of Recent Reaching Experience on Identifying Student Alternate Conceptions

The best possible PCK performance on the whole survey would correspond to selecting the most common incorrect answer choice for each question. This would amount to a score equal to the sum of the largest fractions of incorrect answer choices for each question, which is 9.21. The average FCI-related PCK performance of the 15 instructors who had taught introductory mechanics courses (algebra-based or calculus-based) in the past seven years was 6.33 (standard deviation 0.77), and for the 15 instructors who had not taught them in the past seven years, 6.17 (standard deviation 1.03). There is no statistically significant difference between the two averages ($p$ = 0.634). Thus, recent teaching experience in introductory mechanics did not correlate with the instructors' ability to identify introductory students' alternate conceptions.

**TABLE 1.** Student Alternate Conceptions Related to Newton's Third Law and Identification of Distinct Forces Acting on an Object.

| Introductory Student Alternate Conceptions | FCI Item | % Overall Incorrect | | Incorrect Answer Choices | Introductory Student with Alternate Conceptions | | Instr. |
| --- | --- | --- | --- | --- | --- | --- | --- |
| | | Pre-test | Post-test | | Pre-test | Post-test | |
| Newton's Third Law | | | | | | | |
| If both objects are "active," the larger object exerts the larger force; if only one is active (e.g., car pushing truck), the active object exerts a larger force on the passive object than vice versa | 4 | 74% | 40% | A | 73% | 39% | 97% |
| | 15 | 75% | 56% | C | 61% | 48% | 60% |
| | 16 | 45% | 27% | C | 37% | 19% | 37% |
| Identify forces | | | | | | | |
| Do not know about any forces (including the force of gravity) | 11 | 86% | 65% | E | 3% | 4% | 20% |
| | 29 | 58% | 29% | E | 4% | 1% | 45% |
| Do not know about contact forces (normal force, tension) | 5 | 90% | 76% | A, C, E | 64% | 32% | 70% |
| | 11 | 86% | 65% | A, B, E | 41% | 17% | 60% |
| | 29 | 58% | 29% | A, E | 19% | 3% | 69% |
| Moving object acted on by a distinct force in direction of motion | 5 | 90% | 76% | C, D, E | 86% | 73% | 80% |
| | 11 | 86% | 65% | B, C | 76% | 56% | 63% |

Question scenarios:
Newton's third law
# 4. Truck colliding with car.
# 15. Car pushing truck and speeding up.
# 16. Car pushing truck and moving at constant speed.

Identify all the distinct forces that act on an object
# 5. Identify the forces acting on a ball while moving in a frictionless, circular channel.
# 11. Identify the forces acting on a puck while moving on a frictionless surface.
# 29. Identify the forces acting on a chair at rest on a floor.

## Instructor Identification of Particular Introductory Student Alternate Conceptions

We identified certain student alternate conceptions revealed by FCI questions [1], the questions in which those alternate conceptions are connected to incorrect answer choices, and analyzed the FCI-related PCK performance of instructors on those questions. Similar alternate conceptions were grouped (e.g., those related to Newton's third law, those related to particular tasks such as identifying all the distinct forces acting on an object were placed in a group). If a particular alternate conception appeared in more than one context, we investigated whether instructors performed better at identifying it in some contexts than in other contexts.

For multiple-choice questions, the context comprises both the physical situation in the problem and the answer choices, because different answer choices can change the difficulty of a question. For example, for introductory students a multiple-choice question is easier if the incorrect answer choices are not chosen to reflect common student difficulties, and is challenging when the incorrect answer choices reflect common difficulties [1]. We present two examples of the performance of instructors in identifying alternate conceptions related to Newton's third law and understanding of the distinct forces acting on objects.

Instructors identified student alternate conceptions related to Newton's third law better in some contexts than in others and struggled to identify the most common student alternate conceptions related to forces acting on objects. Here, we discuss the PCK performance of instructors on six questions (short descriptions of the questions are shown in Table 1). Questions 4, 15, and 16 test understanding of Newton's third law, and questions 5, 11, and 29 ask to identify all the distinct forces acting on an object. The average student performance and the PCK performance of instructors are shown in Table 1. The table shows that students hold alternate conceptions related to Newton's third law more strongly in some contexts than in others; thus, student alternate conceptions are context dependent. The FCI-related PCK performance of instructors is also context dependent. For example, virtually all the instructors (97%) identified the alternate conception related to Newton's third law in a typical context (question 4, truck colliding with car), but in other contexts, fewer instructors identified it (e.g., 37% in question 16).

Table 1 also shows that for the three questions that asked students to identify all the distinct forces acting on an object, the majority of instructors were aware that introductory students hold the alternate conception that moving objects are acted on by a distinct force in the direction of motion. However, in all these questions, many instructors claimed that introductory students will not identify contact forces (normal and tension forces) and, to a lesser extent, that they will not identify any forces (including the force of gravity) even in the post-test. However, introductory students rarely selected answer choices that corresponded to these alternate conceptions in the post-test. For example, in question 5, 70% of instructors selected answer choices A, C, and E, which do not include the force that the channel exerts on the ball, selected by (3%, 12%, and 17% of introductory students, respectively). Similarly, in question 11, 60% of instructors selected choices A, B, and E, which do not include the normal force; however, these answer choices combined were selected by only 17% of students. Moreover, in question 29, 45% of instructors predicted that the most common incorrect answer choice selected by introductory students in the post-test is choice E, which states that no forces act on the ball because it is at rest. On the other hand, this answer choice was selected by only 1% of introductory students. Furthermore, 24% of instructors selected choice A (not shown in Table 1), which included only the force of gravity, an answer choice selected by only 1% of introductory students. Thus, instructors had great difficulty identifying introductory students' alternate conceptions related to identification of distinct forces in different contexts. However, with respect to the pre-test data, their performance is better, because before instruction, fewer students were aware of contact forces.

## DISCUSSION AND SUMMARY

Understanding how students reason about physics is important for instructors in designing pedagogical approaches and curricula to account for these difficulties. We found that the performance of instructors on the PCK task was not better for those who had taught introductory mechanics courses recently than for those who had not. Possible reasons for the similarity between the two groups of instructors are that all instructors who taught introductory mechanics employed traditional methods, most had minimal individual contact with students in the large introductory classes, most had not familiarized themselves with physics education literature, and most did not typically grade introductory students' homework and quizzes. Also, even instructors who had not taught introductory algebra- or calculus-based mechanics had taught other introductory courses in which force concepts are relevant, and some had taught these courses more than seven years prior to this study. Many instructors felt that college students in introductory physics classes would not know about contact forces even after instruction. However, the vast majority of college students knew about contact forces. Moreover, instructors' ability to identify introductory students' difficulties with Newton's third law was context dependent.